\documentclass[aps,twocolumn,letterpaper,longbibliography]{revtex4-1}

\usepackage{amssymb}
\usepackage{amsmath}
\usepackage{graphicx}
\usepackage{extarrows}
\usepackage{url}
\usepackage{varioref}
\usepackage{hyperref}
\usepackage{cleveref}

\usepackage{xcolor}
\usepackage[normalem]{ulem}

\begin{document}

\title{Pattern formation in a driven Bose-Einstein Condensate} 

\author{Zhendong Zhang$^1$}
\author{Kai-Xuan Yao$^1$}
\author{Lei Feng$^1$}
\author{Jiazhong Hu$^2$}
\author{Cheng Chin$^1$}

\affiliation{$^1$James Franck Institute, Enrico Fermi Institute and Department of Physics, University of Chicago, Chicago, Illinois 60637, USA}
\affiliation{$^2$Department of Physics and State Key Laboratory of Low Dimensional Quantum Physics, Tsinghua University, Beijing, 100084, China}

\begin{abstract}
    Pattern formation is ubiquitous in nature from morphogenesis
    and cloud formation to galaxy filamentation. More often than not, patterns arise in a medium driven far from equilibrium due to the interplay of dynamical instability and nonlinear wave mixing. We report, based on momentum and real space pattern recognition, 
    formation of density patterns with two- (D$_2$), four- (D$_4$) and six-fold (D$_6$) symmetries in Bose-Einstein condensates (BECs) with atomic interactions driven at two frequencies. The symmetry of the pattern is controlled by the ratio of the frequencies. The D$_6$ density waves, in particular, arise from a resonant wave mixing process that coherently correlates and enhances the excitations that respect the symmetry.
\end{abstract}








\maketitle





How patterns emerge in a homogeneous system is a fundamental question across interdisciplinary research areas including hydrodynamics~\cite{Cross1993}, condensed matter physics~\cite{Mouritsen}, nonlinear optics~\cite{Arecchi}, cosmology~\cite{Liddle} and bio-chemistry~\cite{Turing, Maini}. 
Two paradigmatic examples are Rayleigh-B\'enard convection rolls and Faraday waves ~\cite{Bodenschatz2000,Miles}. Patterns such as stripes, square   lattices and hexagonal lattices form in these systems as a result of wave mixing of different wavelengths
~\cite{Edwards1994,Lifshitz1997,Arbell2002}. A generic model for these phenomena is  described by the Swift-Hohenberg equation~\cite{Swift1977},
\begin{equation}\label{SH}
    \frac{\partial u}{\partial t} = \lambda u - (q^2+\nabla^2)^2 u + f(u),
\end{equation}
where $u=u(\mathbf{x},t)$ is the amplitude of a physical field, $t$ is the evolution time, $q$ determines the momentum of the unstable modes and $\lambda$ characterizes the growth rate at small amplitudes. The nonlinear function $f(u)$ describes the mixing of the modes as the amplitude grows. The shape and the symmetry of the resulting pattern crucially depend on the strength and the form of $f(u)$~\cite{Swift1977,Pomeau1980,Saka1998}.

The onset of pattern formation can be understood from the dynamics and interaction of the excitations in the momentum space, described by the nonlinear amplitude equation \cite{Hoyle2006},
\begin{equation}\label{amp}
    \frac{du_i}{dt} = \alpha_i u_i + \sum\limits_{j,k}\beta_{ijk} u_ju_k + O(u^3),
\end{equation}
where $u_i$ is the amplitude of the $i$-th excitation mode. Starting from small amplitudes, the modes grow exponentially at the rate $\alpha_i$. The quadratic term becomes important as the mode grows, and the tensor $\beta_{ijk}$ describes the mixing of the modes and determines the resulting pattern. The explicit form of $\beta_{ijk}$ is given by the underlying physics, e.g., Navier-Stokes equation for the hydrodynamic systems ~\cite{Landau1959,Teman2001}. 

\begin{figure}[h!]
\centering
\includegraphics[width=86mm]{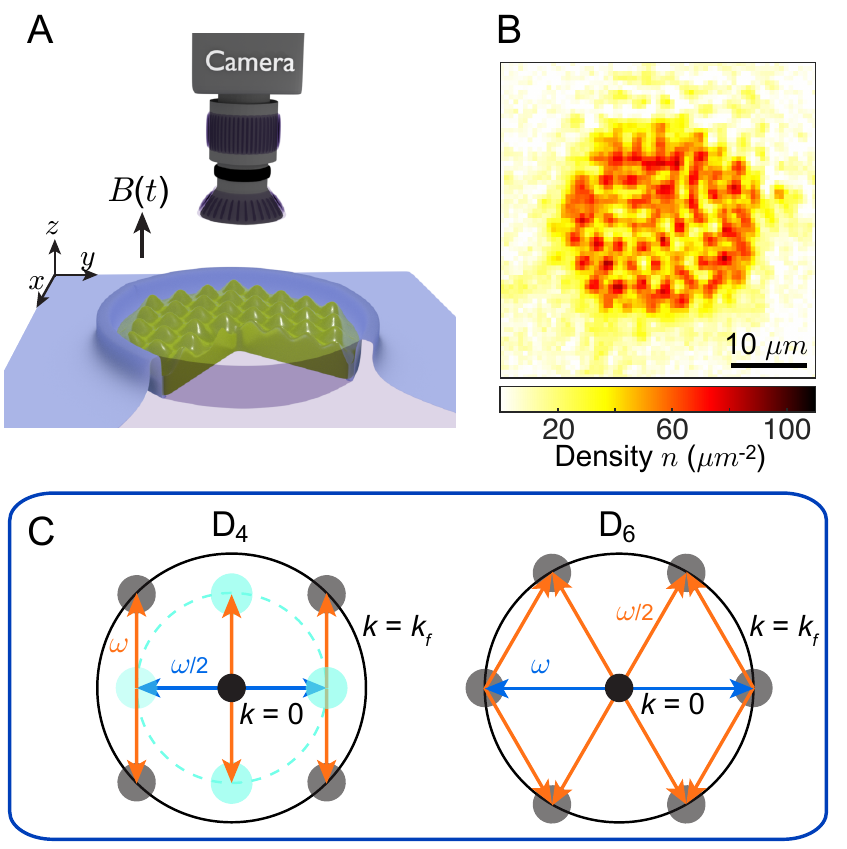}
\caption{\textbf{Pattern formation in a BEC with interaction modulation at two frequencies.}
\textbf{(A)} A BEC (green) of $~^{133}$Cs atoms is trapped in a two dimensional circular potential well (blue). An oscillating magnetic field $B(t)$ in the $z$ direction modulates the scattering length. The atomic density is recorded by a camera.
\textbf{(B)} An example image of the driven BEC displays density waves.
\textbf{(C)} Scattering processes that generate D$_4$ and D$_6$ density waves are illustrated in momentum space in two stages.
In the seeding stage, BEC (black dot) at $k = 0$ produces atom pairs with opposite momentum (blue arrows). In the pattern forming stage, collisions between the atom pairs and the BEC generate four or six modes with $k = k_f$ (orange arrows) with equal angular spacing, which we study in this work.
For the creation of D$_4$ and D$_6$ patterns, the modulation frequencies are $\omega/2$ followed by $\omega$, and $\omega$ followed by $\omega/2$, respectively.
Cyan circles indicate other modes populated during the scattering processes.}
\label{Fig1}
\end{figure}

\begin{figure*}
\centering
\includegraphics[width=172mm]{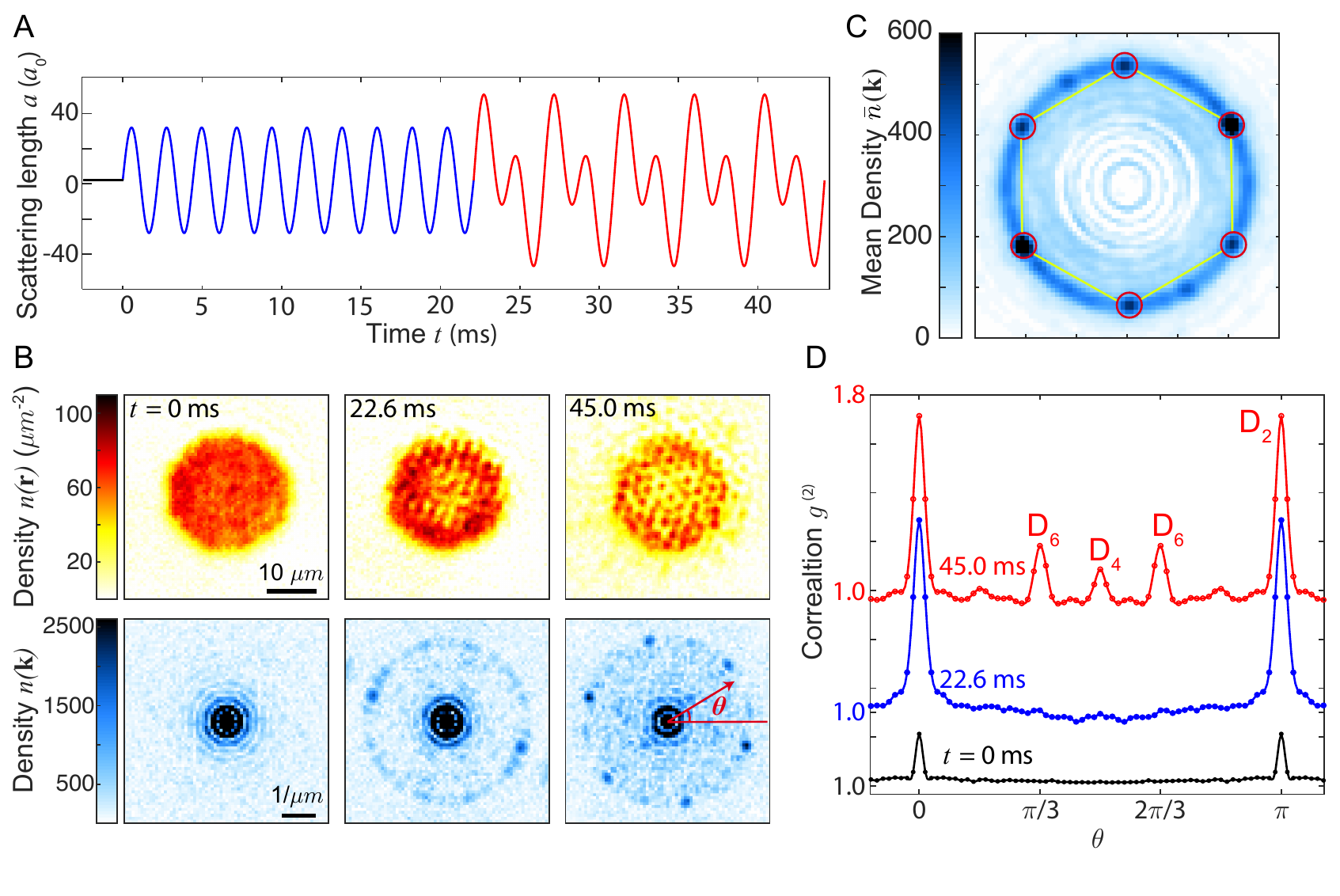}
\caption{\textbf{
Formation of density waves with D$_6$ symmetry.}
\textbf{(A)}The scattering length is modulated in two stages.
The modulation frequency is $ 450$~Hz in the first ten cycles, which is then superposed with a second modulation of $225$~Hz~(see text).
\textbf{(B)} Examples of \textit{in situ} images at times $t = $ 0, 22.6 and 45~ms (top row) and the corresponding Fourier transforms (bottom row). At 45~ms, the Fourier transform displays 6 peaks with $\pi/3$ angular spacing that 
break the rotation symmetry. The 6-peak patterns orient randomly in repeated experiments.  
\textbf{(C)} Pattern recognition based on 185 Fourier transformed images yields six strong peaks (red circles) on the vertices of a hexagon (yellow). Two weaker ones come from patterns with D$_4$ symmetry. We remove the contribution from the BECs ~\cite{Supplement}. 
\textbf{(D)} Correlations $g^{(2)}$ of the Fourier modes with angular spacing $\theta$. The peaks at $\pi/3$, $\pi/2$ and $\pi$ indicate the strength of the patterns with D$_6$, D$_4$ and D$_2$ symmetry respectively.
}
\label{Fig2}
\end{figure*}
In quantum systems, patterns, often characterized by correlation functions, frequently arise from long-range interactions or dynamics far from equilibrium. In polaritonic quantum fluids, hexagonal patterns emerge due to scattering between polaritons \cite{Ardiz2013}. In cold atoms, Faraday waves induced by modulation of trap frequency ~\cite{Engles2007} or interactions~\cite{Nguyen2019} occur in one-dimensional (1D) BECs. BECs also develop spin~\cite{Kron2010} or density wave patterns \cite{Hung2012} by quenches of atomic interaction. Droplets in a dipolar BEC form a hexagonal pattern due to Rosensweig instability \cite{Kadau2016}. Recently, supersolid order, for which a superfluid exhibits spatial correlations, emerges in condensates with spin-orbit coupling \cite{Li2017} or dipolar interactions \cite{Bottcher2019,Tanzi2019,Chomaz2019}. 




In this paper, we report 
formation of various two-dimensional (2D) density wave patterns in a uniform BEC by modulating the atomic interactions at two frequencies (Fig.~\ref{Fig1}A). The interaction modulation is realized by applying an  oscillating magnetic field to the sample~\cite{Clark2017,Fu2018}. The magnetic field is in the $z-$direction, perpendicular to the sample while the pattern forms in the horizontal $x-
y$ plane (Fig.~\ref{Fig1}B). By changing the ratio of the two modulation frequencies, density patterns with D$_2$,  D$_4$ and D$_6$ symmetries are observed \textit{in situ} and analyzed. The D$_6$ density wave pattern, in particular, results from a novel coherent process that resonantly couples six momentum modes.

To understand the pattern formation process in a driven condensate, we derive the associated quantum nonlinear amplitude equation as~\cite{Supplement}

\begin{equation}
    \frac{d\hat{a}_{\mathbf{k}}}{dt} = \gamma_1 \hat{a}^\dagger_{-\mathbf{k}} + \gamma_2\sum\limits_{\mathbf{k}_1}\hat{a}^\dagger_{\mathbf{k}_1-\mathbf{k}}\hat{a}_{\mathbf{k}_1}-\gamma_2^*\sum\limits_{\mathbf{k}_2}\hat{a}_{\mathbf{k}_2}\hat{a}_{\mathbf{k}-\mathbf{k}_2},
    \label{Eq_of_motion}
\end{equation}
where $\hat{a}_{\mathbf{k}}$ and $\hat{a}^\dagger_{\mathbf{k}}$ are the bosonic annihilation and creation operators with momentum $\hbar\mathbf{k}$, respectively, $\hbar = h/2\pi$ is the reduced Planck constant, the summations include all resonant scattering processes, and the rate constants $\gamma_1$ and $\gamma_2$ are given by the modulation strengths. This equation is reminiscent of the classical amplitude equation Eq.~\ref{amp}. 


The wave mixing processes leading to D$_4$ and D$_6$ patterns can be described in two stages (Fig.~\ref{Fig1}C). In the \textit{seeding} stage, atom pairs with opposite momentum are generated from the condensate by a single-frequency modulation. Such a process, given by the first term in Eq.~\ref{Eq_of_motion}, seeds and amplifies the primary excitations that spontaneously break the rotational symmetry of the system. In the \textit{pattern forming} stage, the same or a different frequency component is introduced to the modulation, which stimulates scatterings into a particular pattern with the desired symmetry~\cite{Supplement}. This process is described by the nonlinear terms in Eq.~\ref{Eq_of_motion}. Finally, the excitation modes interfere with the BEC to form the density wave $n(r)$, which we observe. The density wave relates to the excitations $\hat{a}_\mathbf{k}$ as
$\hat{n}(r) = n_0[\hat{1} +  N_0^{-1/2}\sum\limits_\mathbf{k} (\hat{a}_\mathbf{k}+\hat{a}_{-\mathbf{k}}^\dagger) e^{i\mathbf{k}\cdot r}]$, where $n_0$ is the condensate density and $N_0\gg1$ is the atom number in the condensate.

The experiment starts with a BEC of $N_0 = 60,000$ cesium atoms in a dipole trap. Atoms are radially confined in a circular potential well with a radius of 14.5~$\mu$m and a barrier height of $h\times 140$~$\mathrm{Hz}$. In the vertical direction, the sample is confined in a harmonic potential with a $1/e^2$ radius of 0.78~$\mu$m.
We then apply an oscillating magnetic field near a Feshbach resonance \cite{Chin2010} to the BEC, which modulates the atomic $s$-wave scattering length $a$. After modulation time $t$, we perform \textit{in situ} imaging to record the density waves.

\begin{figure*}
\centering
\includegraphics[width=140mm]{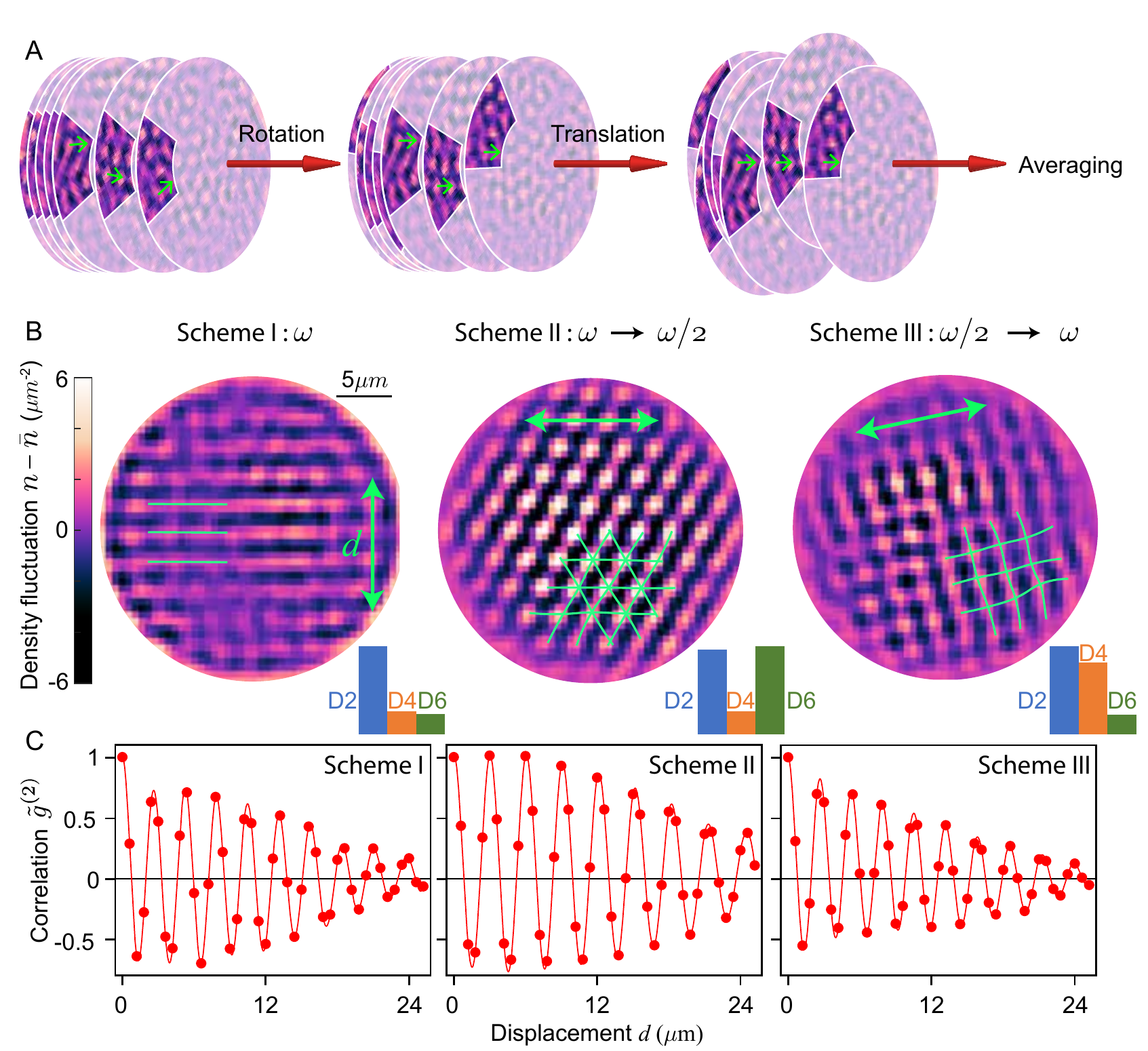}
\caption{\textbf{Density wave patterns in real space.}
\textbf{(A)} In our pattern recognition algorithm, each of the \textit{in situ} images is first rotated and then translated to overlap the density waves. The translation maximizes the variance of the averaged image~\cite{Supplement}. \textbf{(B)} Resulting density waves from the algorithm for Scheme I (stripes): single modulation frequency $\omega$, Scheme II (hexagonal lattice): $\omega$ followed by $\omega/2$, and Scheme III (square lattice): $\omega/2$ followed by $\omega$. 
The green lines are guides to the eye to highlight the corresponding pattern. The green arrows show the direction along which the real space correlation is evaluated in panel C. The bar diagrams show the relative weights of D$_2$, D$_4$ and D$_6$ symmetry components from fitting the patterns~\cite{Supplement}.  
\textbf{(C)} Real space correlation functions evaluated from the patterns. The oscillations have periods of 2.63(1), 3.05(1) and 2.65(1)~$\mu$m for schemes I, II and III, respectively. The ratio of the periods is $1.156(2)$, consistent with theory value $2/\sqrt{3} \approx 1.155$. The solid lines are guides to the eye.}
\label{Fig3}
\end{figure*}

\begin{figure*}
    \centering
    \includegraphics[width = 172mm]{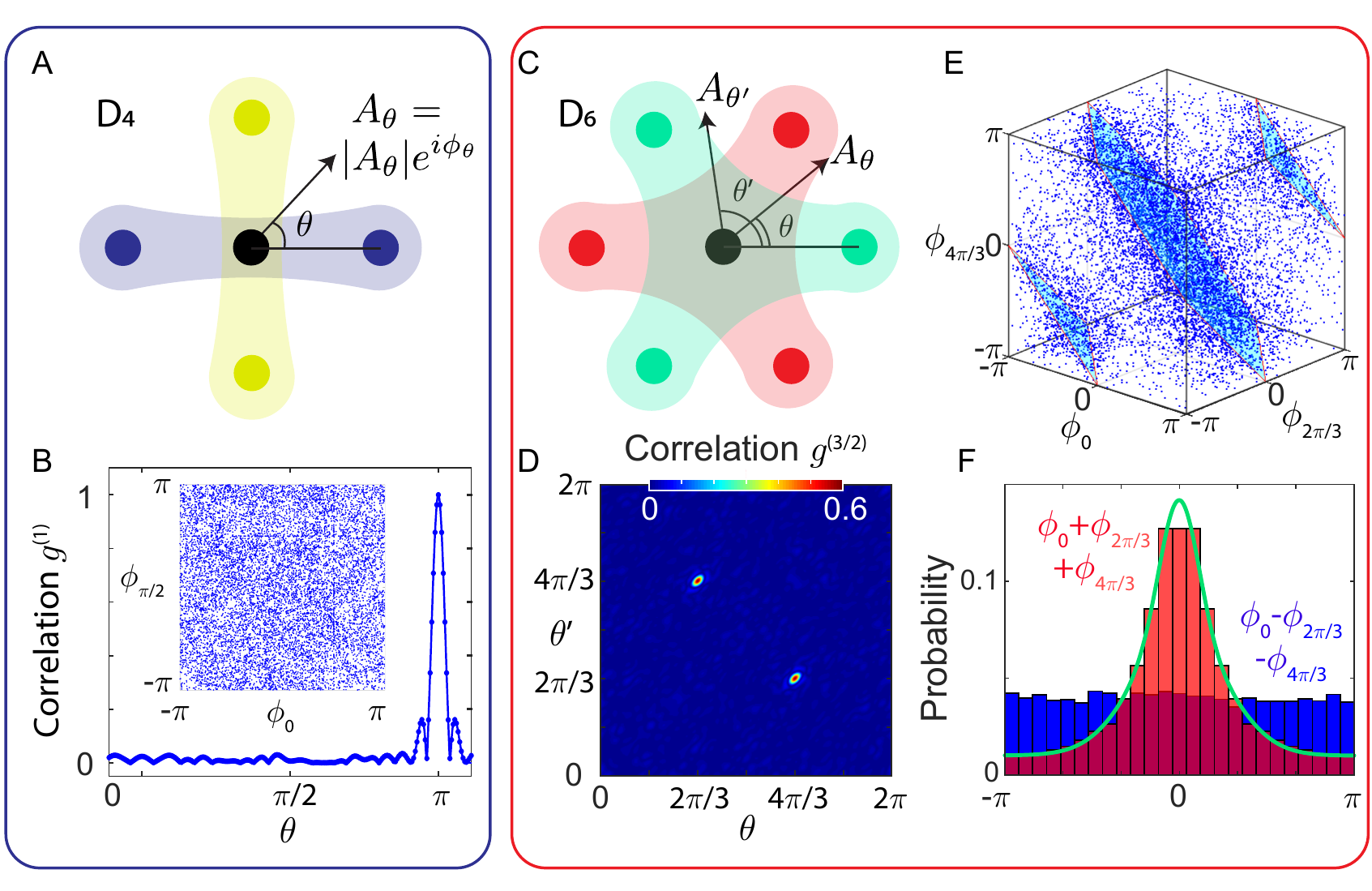}
\caption{\textbf{Coherent properties of D$_4$ and D$_6$ density waves.}
\textbf{(A)} Pairs of modes with opposite momenta (blue and yellow) are phase correlated in the $\mathrm{D}_4$ density wave pattern. 
\textbf{(B)} Phase correlation function $g^{(1)}$ between Fourier components of  the density wave is evaluated based on 123 images obtained with Scheme III. Strong  correlations appear between modes moving in opposite directions $\theta = \pi$, but not perpendicular modes $\theta = \pi/2$. The inset is the histogram of the phase of perpendicular modes $\pi/2$, which shows no discernable correlation. \textbf{(C)} Three-point phase coherence appears in triplet modes (green and red) of the  $\mathrm{D}_6$ density wave pattern. \textbf{(D)} Phase correlation function $g^{(3/2)}$ of three Fourier amplitudes separated by angles $\theta$ and $\theta'$, evaluated based on 185 images with Scheme III, shows two peaks at $(\theta=2\pi/3, \theta'=4\pi/3)$ and $(4\pi/3, 2\pi/3)$, supporting phase correlations of the triplets. 
\textbf{(E)} Phases of three modes separated by $2\pi/3$ and $4\pi/3$ show higher probability near the planes $\phi_{0}+\phi_{2\pi/3}+\phi_{4\pi/3}=0, \pm2\pi$ (blue planes).
\textbf{(F)} The probability distribution (red) of the phase $\phi_{0}+\phi_{2\pi/3}+\phi_{4\pi/3}$ weighted by the atom number of the triplet modes displays a peak at $0$ (red bars). An Alternative combination of the phases, $\phi_{0}-\phi_{2\pi/3}-\phi_{4\pi/3}$, is evenly distributed (blue bars). The green curve is from the numerical calculation~\cite{Supplement}. }
\label{Fig4}
\end{figure*}

We first describe the experimental procedure for the formation of the density waves with D$_6$ symmetry. In the seeding stage, we apply a single -frequency modulation as $a(t) = a_{dc} + a_1\sin{\omega t}$, where $\omega/2\pi=450$~Hz, $a_1 = 30~a_0$, $a_{dc} = 2~a_0$ and $a_0$ is the Bohr radius. After $t = 22.2$~ms, in the pattern forming stage, we add a second frequency component to the modulation as $a(t) = a_{dc}+a_1\sin\omega t+a_2\sin\omega t/2$, where $a_2=25~a_0$~(see Fig.~\ref{Fig2}A).

We analyze the symmetry of the density wave patterns based on Fourier analysis. In the seeding stage, only stripe patterns appear. In the pattern forming stage, hexagonal lattice patterns with D$_6$ symmetry emerge, signified by six distinct modes in the Fourier space. The modes are equally spaced by $\pi/3$ in their directions with the same wavenumber $k_f = \sqrt{m\omega/\hbar}$ ~\cite{Clark2017} (see Fig.~\ref{Fig2}B), where $m$ is the atomic mass. 

The presence of the D$_6$ pattern can be further confirmed with a pattern recognition algorithm \cite{Feng2019} (see Fig.~\ref{Fig2}C). 
To quantify the strength of the patterns, we evaluate the density correlation function $g^{(2)}(\theta) \equiv \langle |A_\varphi|^2|A_{\varphi+\theta}|^2\rangle/\langle |A_\varphi|^2 \rangle^2$,
where $A_\theta = \int n(\mathbf{r})e^{-i\mathbf{k}_\theta\cdot\mathbf{r}}d\mathbf{r}$
is the Fourier amplitude evaluated at $\mathbf{k}_\theta$ with magnitude $|\mathbf{k}_\theta|=k_f$ and angle $\theta$. The angle brackets denote averaging over both the angle $\varphi$ from 0 to $2\pi$ and the images. The evolution of $g^{(2)}$ confirms the growth of different patterns in the seeding and pattern forming stages (Fig.~\ref{Fig2}D).

We tailor the modulation waveform  to create different patterns.
Here three modulation schemes that lead to patterns with D$_2$, D$_4$ and D$_6$ symmetries are reported. Scheme I: we apply the modulation at a single frequency $\omega$. Scheme II: we modulate at frequency $\omega$ in the seeding stage and superpose a second frequency $\omega/2$ in the pattern forming stage (Fig.~\ref{Fig2}A). Scheme III: we modulate at frequency $\omega/2$ and then switch to frequency $\omega$.


To reveal the density patterns in real space, we employ a 2D pattern recognition algorithm. 
Since the pattern in each image appears with random orientation and displacement, 
the algorithm is developed to rotate and align the patterns (Fig.~\ref{Fig3}A). 
We determine the orientation of each image as illustrated in Fig.~\ref{Fig2}C, and align all of them in the same direction. We then translate each of the images independently to maximize the spatial variance of their average. Finally we extract the underlying pattern by averaging all aligned images. To eliminate long wavelength variations that are uncorrelated with the pattern, we filter the density fluctuations at $|\mathbf{k}| \leq 0.75k_f$ from the images to get the density waves $\tilde{n}(\mathbf{r})$.

The results of the 2D pattern recognition algorithm are shown in Fig.~\ref{Fig3}B. Single frequency modulation (Scheme I) produces D2 stripe patterns. Scheme II ($\omega \rightarrow \omega/2$) results in a hexagonal lattice pattern, consistent with Fig.~\ref{Fig2}. Scheme III ($\omega/2 \rightarrow \omega$) results in a square lattice pattern. We further determine the strengths of different symmetry components in each image $P$ based on the fit: $P = c_2 P_2+c_4 P_4+c_6 P_6$,
where $P_n$ are normalized patterns with D$_n$ symmetry, and $c_n$ are the fitting parameters ~\cite{Supplement}. The results, shown in the bar diagrams of Fig.~\ref{Fig3}B, suggest that different schemes are effective in generating patterns with different symmetries.



Remarkably, all three patterns extend throughout the entire sample. The spatial extent of the patterns can be evaluated from their real space correlation functions $\tilde{g}^{(2)} (\mathbf{r}) \equiv \int \tilde{n}(\mathbf{r}_0) \tilde{n}(\mathbf{r}_0 + \mathbf{r}) d\mathbf{r}_0/\int \tilde{n}(\mathbf{r}_0)^2 d\mathbf{r}_0$.
Correlations along principle directions, shown in Fig.~\ref{Fig3}C, extend across the entire sample of diameter 25~$\mu$m. Comparing the patterns, we observe that the D$_6$ pattern is a factor of 5 more pronounced than D$_4$ even though these two schemes employ similar modulation strengths~\cite{Supplement}.

The clear difference between the strength of the D$_4$ and D$_6$ patterns  comes from the coherence of the underlying scattering processes. For D$_4$ patterns, phase coherence only exists between counter-propagating modes, illustrated in Fig.~\ref{Fig4}A. We evaluate the two-point phase correlation function of the density waves as $g^{(1)}(\theta) \equiv \langle A_\varphi A_{\varphi+\theta}\rangle/\langle|A_\varphi|^2\rangle$,
where $A_\theta = |A_\theta| e^{i \phi_\theta}$ is the Fourier amplitude of the mode with wavenumber $k_f$ at angle $\theta$ and $\phi_\theta$ is its phase.
The result, see Fig.~\ref{Fig4}B, shows a single peak at $\theta = \pi$, simply due to the realness of density. The absence of other features, particularly at $\theta=\pi/2$, shows that the density waves in orthogonal directions are incoherent. Close inspection of the phases of orthogonal modes, see inset of Fig.~\ref{Fig4}B, confirms the absence of correlation. 

The D$_6$ pattern, on the other hand, displays a novel phase coherence in triplets of modes angularly spaced by 2$\pi$/3, see Fig.~\ref{Fig4}C. Here we evaluate the three-point phase correlation function as 
\begin{equation}
    g^{(3/2)} (\theta,\theta') \equiv \frac{\langle A_\varphi A_{\varphi+\theta}A_{\varphi+\theta'}\rangle}{\sqrt{\langle|A_\varphi|^2\rangle\langle|A_{\varphi+\theta}|^2\rangle\langle|A_{\varphi+\theta'}|^2\rangle}}.
    \label{g_32_main}
\end{equation}
The correlation shows two peaks at $(\theta,\theta') = (2\pi/3, 4\pi/3)$ and $(4\pi/3,2\pi/3)$ (see Fig.~\ref{Fig4}D), where $\theta$ and $\theta'$ are the relative angles between the three modes. This indicates phase coherence of any three modes angularly separated by $2\pi/3$. From repeated measurements, we find that the phases of the triplets are statistically constrained to $\phi_{0}+\phi_{2\pi/3}+\phi_{4\pi/3} = 0$ modulo $2\pi$ with a small standard deviation of $\delta\phi=1.1$, see Fig.~\ref{Fig4}E and F. The phase differences, e.g. $\phi_{0}-\phi_{2\pi/3}-\phi_{4\pi/3}$, as well as other permutations, are uniformly distributed and thus uncorrelated.

The three-point phase correlation is an essential element to understanding the growth and the origin of D$_6$ patterns in our system. Based on Eq.~(3), we show that the strength of the $\mathrm{D}_6$ pattern satisfies the equation of motion~\cite{Supplement}

\begin{equation}
    \frac{dA_{rms}}{dt}=\gamma_1 A_{rms}+\gamma_2 g^{(3/2)} A_{rms}^2,
\end{equation}

\noindent where $A_{rms}$ is the root-mean-square of the six Fourier amplitudes that constitute the $\mathrm{D}_6$ pattern and $g^{(3/2)}\equiv g^{(3/2)}(2\pi/3,4\pi/3)$. A positive $g^{(3/2)}$ suggests that beyond small amplitudes, the nonlinear wave mixing term dominates and leads to a faster-than-exponential (hyperbolic) growth of the D$_6$ density waves. The large measured value of $g^{(3/2)} = 0.58$ explains the strong D$_6$ pattern that we observe. 

How does the three-point phase correlation emerge in a driven condensate? Starting from a condensate seeded by the single-frequency modulation, we see that $g^{(3/2)}$ increases quickly from zero after the two-frequency modulation starts~\cite{Supplement}. Theoretically the growth of the correlation is linked to the resonant nonlinear coupling of excitation modes that respect the symmetry and is described by $dg^{(3/2)}/dt=3\gamma_2 A_{rms}$ for small amplitudes $A_{rms}<<N_0^{1/2}$. Our measurement is in good agreement with the theory~\cite{Supplement}. Given the above, the three-point phase relation $\phi_{0}+\phi_{2\pi/3}+\phi_{4\pi/3} = 0$ (see Fig.~4F) can be understood as the phase matching condition that maximizes the correlator $g^{(3/2)}$, which explains the dominance of the D$_6$ pattern in our experiment.\\
\\
\noindent\textbf{Acknowledgement}

\noindent We thank K. Patel for careful reading the manuscript. This work is supported by National Science Foundation (NSF) grant no. PHY-1511696, the Army Research Office Multidisciplinary Research Initiative under grant W911NF-14-1-0003 and the University of Chicago Materials Research Science and Engineering Center, which is funded by the NSF under grant no. DMR-1420709. J.H. acknowledges the financial support from National Natural Science Foundation of China under grant no. 11974202.


\clearpage
\widetext
\setcounter{equation}{0}
\setcounter{figure}{0}
\setcounter{table}{0}
\setcounter{page}{1}
\makeatletter
\renewcommand{\theequation}{S\arabic{equation}}
\renewcommand{\thefigure}{S\arabic{figure}}
\renewcommand{\thetable}{S\arabic{table}}

\noindent \textbf{Materials and Methods}\\

\section{Experimental Procedure}
We start with BECs of 60,000 cesium atoms loaded into a disk-shaped dipole trap with a radius of $14.5~\mu m$ in the horizontal direction. The horizontal confinement is provided by a blue-detuned laser at 780~nm. We shape the laser beam profile using a digital micromirror device and project it to the atom plane through a high-resolution objective. The resulting circular potential well has a barrier height of $h\times 140~\mathrm{Hz}$. Atoms are also tightly confined in the vertical direction with a $1/e^2$ radius of $0.78~\mu m$ with a harmonic trapping frequency of $259~\mathrm{Hz}$.

After preparing the sample, we modulate the magnetic field near a Feshbach resonance, which causes the s-wave scattering length $a$ of the atoms to oscillate as $a(t) = a_{dc} + a_1(t)\sin{\omega_1t} + a_2(t)\sin{(\omega_2t+\phi)}$. We edit the control voltage output from an arbitrary waveform generator to modulate currents in the coils, which leads to the magnetic field being modulated according to a designed waveform. A small positive offset scattering length $a_{dc} = 2a_0$ is maintained throughout the experiment to keep the condensate stable. For generating the $\mathrm{D}_2$ density wave pattern, we keep modulating the scattering length at frequency $450~\mathrm{Hz}$ with amplitude $45~a_0$ for 23.8 ms. For $\mathrm{D}_4$ pattern, we first modulate at $225~\mathrm{Hz}$ for 3 cycles with amplitude $45~a_0$ and then switch to $450~\mathrm{Hz}$ with the same amplitude for 24 ms. To generate $\mathrm{D}_6$ pattern, the first 10 cycles of modulation is at $450~\mathrm{Hz}$ with amplitude $30~a_0$, which is then mixed with another frequency component at $225~\mathrm{Hz}$ and amplitude $25~a_0$ for $22.8~\mathrm{ms}$. The relative phase $\phi$ between these two frequency components is 0.

We finally perform \textit{in situ} absorption imaging to observe the resulting density waves in condensates using the high-resolution objective and a CCD camera. Our imaging system is sensitive to density fluctuations of spatial frequency ranging from 0 up to $3.44~\mu m^{-1}$\cite{Hung2011}, which covers  the density waves we observe at $k_f = 2.43~\mu m^{-1}$. The individual pixel size of the CCD camera is $0.6~\mu m$, which provides a sampling frequency of 4 data points within one wavelength of the density waves.

In order to extract the population of excited modes from their interference with the condensate, we first Fourier transform the images including density waves with $121\times121$ pixels. Then in the Fourier space we focus on the ring at $|k - k_f|\leq0.1k_f$ and cut it using angular slices of $3^\circ$ to count the average Fourier magnitude $A_\theta$ in the direction at angle $\theta$. In general, the sensitivity of our imaging system varies for signals with different wavenumber. We measure the modulation transfer function $M(\mathbf{k})$ of thermal atoms and find that the proportional constant of measured strength of density fluctuations at $k_f$ to its corresponding real strength is $M(k = k_f) = 0.45$~\cite{Hung2011}. The relation between density wave amplitude $A_\theta$ and population $|a_{\mathbf{k}}|^2$ is $|A_\theta|^2 = 4N_0\cos^2{(\omega t/2)}|a_{\mathbf{k}}|^2$, where the phase $\omega t/2\approx0.57 ~\mathrm{rad}$ at the time we perform the imaging. Finally the population is evaluated as $|a_{\mathbf{k}}|^2 = |A_\theta|^2/[M^2(k=k_f)4N_0cos^2(\omega t/2)]$. Also, we observe the density waves stroboscopically every 4.4~ms as shown in Fig.~\ref{Fig_S4}.

\section{Quantum dynamics of pattern formation}
We start from the general form of Hamiltonian of driven BECs,
\begin{align}
    H = \int d^3\mathbf{r}\Psi^\dagger(\mathbf{r},t)\frac{p^2}{2m}\Psi(\mathbf{r},t) + \int d^3\mathbf{r} \Psi^\dagger(\mathbf{r},t)V(\mathbf{r})\Psi(\mathbf{r},t)+\frac{g(t)}{2}\int d^3\mathbf{r}\Psi^\dagger(\mathbf{r},t)\Psi^\dagger(\mathbf{r},t)\Psi(\mathbf{r},t)\Psi(\mathbf{r},t),
    \label{H}
\end{align}
where the interaction strength is modulated as $g(t) = \frac{4\pi\hbar^2}{m}[a_{dc}+a_1(t)\sin{\omega_1 t}+a_2(t)\sin(\omega_2 t+\phi)]$. Here $a_{dc}$ is a small offset scattering length to keep the condensate stable, $a_{1,2}$ are amplitudes of scattering length modulation and $\phi$ is the relative phase between the two frequency components $\omega_1$ and $\omega_2$.

The external potential $V(\mathbf{r})$ is neglected later because it only serves to determine the initial wavefunction of BECs and doesn't affect the dynamics. After doing the Fourier transform $\Psi(\mathbf{r}) = \frac{1}{\sqrt{V}}\sum_{\mathbf{k}}\hat{a}_{\mathbf{k}}e^{i\mathbf{k}\cdot\mathbf{r}}$, we obtain the Hamiltonian in momentum space as
\begin{align}
    H = \sum_{\mathbf{k}}\epsilon_{\mathbf{k}}\hat{a}^\dagger_{\mathbf{k}}a_{\mathbf{k}} + \frac{g(t)}{2V}\sum_{\mathbf{k}_1,\mathbf{k}_2,\Delta\mathbf{k}}\hat{a}^\dagger_{\mathbf{k}_1+\Delta\mathbf{k}}\hat{a}^\dagger_{\mathbf{k}_2-\Delta\mathbf{k}}\hat{a}_{\mathbf{k}_1}\hat{a}_{\mathbf{k}_2},
    \label{H_k}
\end{align}
where $V$ is the volume of condensate and the dispersion is $\epsilon_{\mathbf{k}} = \hbar^2k^2/2m$.

After transferring to the rotating frame with $\hat{a}_{\mathbf{k}}\rightarrow\hat{a}_{\mathbf{k}}e^{-i\epsilon_{\mathbf{k}}t/\hbar}$ and using the rotating wave approximation to eliminate the fast oscillating terms, the Hamiltonian becomes time-independent:

\begin{align}
    H_I = \frac{i}{4V}(\sum_{\mathbf{k}}g_1\hat{a}_{\mathbf{k}}^\dagger\hat{a}_{-\mathbf{k}}^\dagger\hat{a}_0\hat{a}_0 + \sum_{\mathbf{k}'}g_2\hat{a}_{\mathbf{k}'}^\dagger\hat{a}_{-\mathbf{k}'}^\dagger\hat{a}_0\hat{a}_0+\sum_{\mathbf{k}_1,\mathbf{k}_2}e^{-i\phi}\hat{a}_{\mathbf{k}_2}^\dagger\hat{a}_{\mathbf{k}_1-\mathbf{k}_2}^\dagger\hat{a}_{\mathbf{k}_1}\hat{a}_0) + h.c.,
\end{align}
where $g_1 = 4\pi\hbar^2a_1/m$ and $g_2 = 4\pi\hbar^2a_2/m$ and the summations go over the processes that satisfy the following energy conservation conditions:
\begin{align}
\epsilon_{\mathbf{k}} + \epsilon_{-\mathbf{k}} &= \hbar\omega_1\nonumber,\\
\epsilon_{\mathbf{k}'} + \epsilon_{-\mathbf{k}'} &= \hbar\omega_2\nonumber,\\
\epsilon_{\mathbf{k}_2} + \epsilon_{\mathbf{k}_1-\mathbf{k}_2}
 &=
\epsilon_{\mathbf{k}_1} + \hbar\omega_2.
\end{align}
Here the left/right hand side is the total energy after/before the collision.

Then the equation of motion for $\hat{a}_{\mathbf{k}}$ is obtained to second order in the Bogoliubov approximation $\hat{a}_0 \approx \hat{a}^\dagger_0\approx\sqrt{N_0}$ as
\begin{align}
\frac{d\hat{a}_\mathbf{k}}{dt} = \gamma_1\hat{a}^\dagger_{-\mathbf{k}} + \gamma_2\sum_{\mathbf{k}_1}\hat{a}^\dagger_{\mathbf{k}_1-\mathbf{k}}\hat{a}_{\mathbf{k}_1} - \gamma_2^*\sum_{\mathbf{k}_2}\hat{a}_{\mathbf{k}_2}\hat{a}_{\mathbf{k}-\mathbf{k}_2},
\label{Eq4}
\end{align}
where the growth rates are given by $\gamma_1 = \frac{N_0\pi\hbar a_1}{mV}$, $\gamma_2 = \frac{\sqrt{N_0}\pi\hbar a_2}{mV}e^{-i\phi}$. Here all the momenta are restricted to the horizontal plane and the magnitude of $\mathbf{k}$ is $|\mathbf{k}| = k_f = \sqrt{m\omega_{1}/\hbar}$. We have been using $\omega_1/2\pi = 450~\mathrm{Hz}$ and $\omega_2/2\pi = 225~\mathrm{Hz}$.  

The formation of density wave patterns originates from the momentum and energy conservation of underlying bosonic stimulated scattering processes (see Fig.~1C). For the $\mathrm{D}_4$ pattern formation under Scheme III, during the modulation of frequency $\omega_2$, a pair of BEC atoms absorb a quantum of energy $\hbar\omega_2$ and scatter into a pair of atoms with opposite momenta $\pm \mathbf{k}_{1}$ at $|\mathbf{k}_1| = k_f/\sqrt{2}$ and energy $\epsilon_{\mathbf{k}_1} = \hbar\omega_2/2$. Then one atom with $\mathbf{k}_1$ collides with one BEC atom absorbing another quantum of $\hbar\omega_2$. One of them scatters into $\mathbf{k}$ with magnitude $k_f$ and energy $\epsilon_{\mathbf{k}} = \hbar\omega_2$ at $45^\circ$(or $-45^\circ$) relative to $\mathbf{k}_1$. The other one is scattered into $\mathbf{k}_1-\mathbf{k}$ with magnitude $k_f/\sqrt{2}$ and energy $\epsilon_{\mathbf{k}_1-\mathbf{k}} = \hbar\omega_2/2$ at $-90^\circ$ (or $90^\circ$) relative to $\mathbf{k}_1$. This process is described by the second term on the right hand side (RHS) of Eq.~\ref{Eq4}. On the other hand, one atom with $-\mathbf{k}_1$ can collide with one BEC atom and one of the scattered atoms has momentum $k_f$ at $45^\circ$ or $-45^\circ$ relative to $-\mathbf{k}_1$. Thus, seeds of 4 momentum modes at $k_f$ with $90^\circ$ relative angular spacing are generated. Later, when another modulation of frequency $\omega_1 = 2\omega_2$ is applied, those 4 modes get amplified with pairs of BEC atoms scattering into them. This corresponds to the first term on the RHS of Eq.~\ref{Eq4}. Finally the those 4 momentum modes with $90^\circ$ angular spacing interfere with the BEC to form the $\mathrm{D}_4$ density wave pattern.

On the other hand, for $\mathrm{D}_6$ pattern formation under Scheme II, a modulation of frequency $\omega_1$ is first applied to generate pairs of opposite momentum modes $\pm\mathbf{k}$ at $k_f$ and energy $\epsilon_{\mathbf{k}} = \hbar\omega_1/2$. Then when the second frequency component $\omega_2 = \omega_1/2$ is added, an atom with $\mathbf{k}$ collides with a BEC atom absorbing one energy quantum $\hbar\omega_2$ and scattering into atoms with $\mathbf{k}_2$ and $\mathbf{k}-\mathbf{k}_2$ with the same magnitude $k_f$ and energy $\epsilon_{\mathbf{k}_2} = \epsilon_{\mathbf{k}-\mathbf{k}_2} = \hbar\omega_1/2$ at $\pm 60^\circ$ relative to $\mathbf{k}$. This corresponds to the third term on the RHS of Eq.~\ref{Eq4}. Also, atoms with momentum $\mathbf{k}_2$ or $\mathbf{k}-\mathbf{k}_2$ can collide with one BEC atom into atoms with $\mathbf{k}$, corresponding to the second term on the RHS of Eq.~\ref{Eq4}. Also, one atom with $-\mathbf{k}$ can collide with one BEC atom and scatter into $-\mathbf{k}_2$ or $-(\mathbf{k}-\mathbf{k}_2)$ at $\pm 60^\circ$ relative to $-\mathbf{k}$. Thus, 6 momentum modes with $60^\circ$ relative angular spacing are generated and are amplified by the $\omega_1$ frequency component at the same time. Eventually they interfere with the condensate and form the $\mathrm{D}_6$ density wave pattern.

\section{Principal component analysis}

In order to remove the background of Fourier space in Fig.~\ref{Fig2}C, we collect 100 images of pure BECs and apply PCA algorithm to construct the bases and subtract the projection onto these bases from the Fourier transform of BECs with density waves. 

We first get the Fourier amplitude's magnitude $n_i(\mathbf{k})$ of the $i^{th}$ image of pure BEC atomic density $n_i(\mathbf{r})$. Each $p\times p$ square matrix $n_i(\mathbf{k})$ is rearranged into a $1\times p^2$ row vector. Then all the row vectors are arranged to form a rectangular matrix $M_{ij}$, where $j$ ranges from $1$ to $p^2 = 121^2$. The mean value of each column is shifted to zero by subtracting the average of experimental realizations, resulting in the data matrix $X = M - \Bar{M}$. Our goal is to diagonalize the covariance matrix $X^{T}X$ to find its eigenvectors $w_j$ and eigenvalues $\lambda_j$, which corresponds to statistical independent bases (principal components) and variance of $X$'s projection $X_{ij}w_j$ onto each basis, respectively. We use singular value decomposition (SVD) to perform this diagonalization.

\begin{figure}
    \centering
    \includegraphics[width = 86mm]{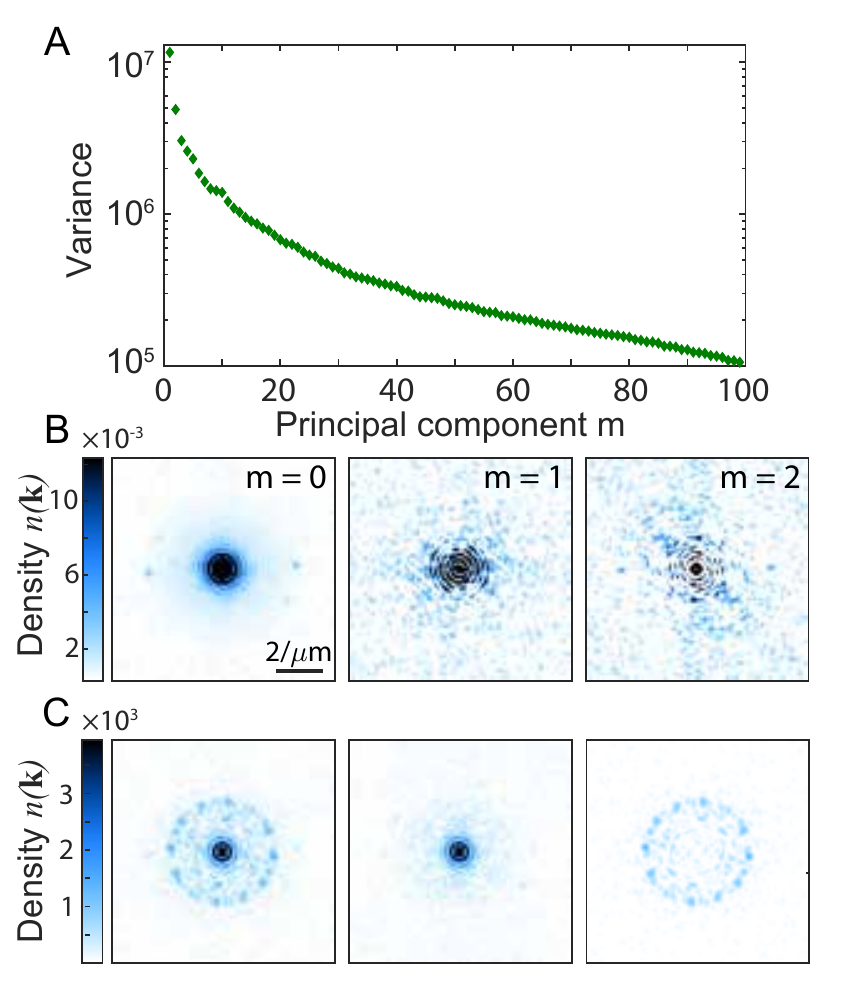}
    \caption{\textbf{Principal component analysis for removing the background in Fourier space.} \textbf{(A)} The variance of the data matrix $X$'s projection onto each principal components. The y axis is in log scale. \textbf{(B)} The mean of the magnitude of Fourier transforms of pure BECs (left) and the first two bases from PCA (middle and right). \textbf{(C)} One example of removing all the projection onto PCA bases (middle) from the Fourier space of BEC with density waves (left), only signals from the density waves survive (right).}
    \label{FigS1}
\end{figure}

The first 99 principal components are kept and the corresponding variances are shown in Fig.~\ref{FigS1}A. The average of $n_i(\mathbf{k})$ is counted as an additional basis $w_0$. In Fig.~\ref{FigS1}B, we plot the average of $n_i(\mathbf{k})$ and the two principal components that have the largest and second largest variances. Next, we use those constructed bases to remove the background in the Fourier space $n_d(\mathbf{k})$ of the atomic densities of BECs with density waves $n_d(\mathbf{r})$. As an example, in Fig.~\ref{FigS1}C, we project one $n_d(\mathbf{k})$ to all the principal components $w_j$ to reconstruct the background. Finally the background is subtracted from the original Fourier space and only the signals from density waves are left.

\begin{figure*}
    \centering
    \includegraphics[width = 172mm]{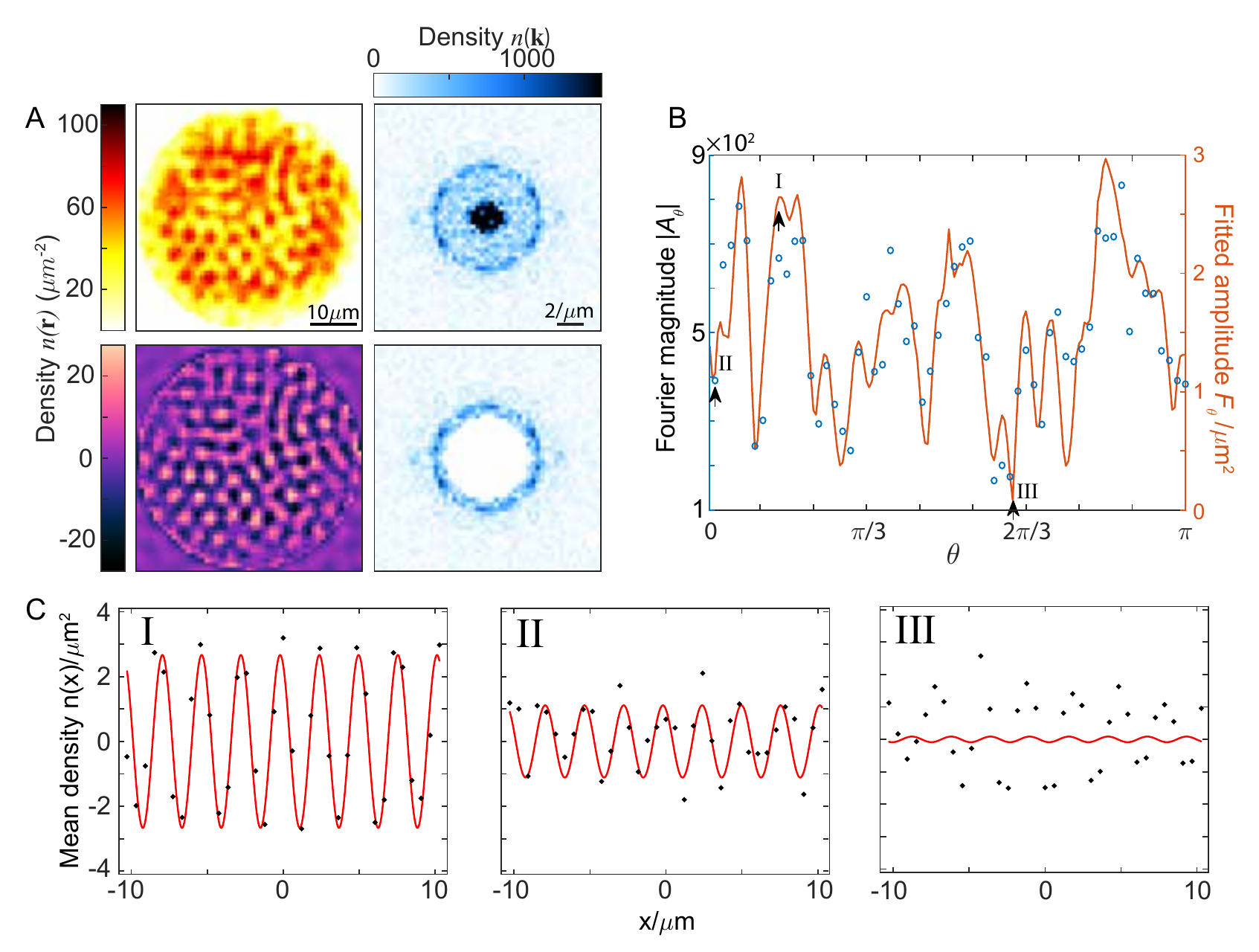}
    \caption{\textbf{Extraction of the phases and amplitudes of density waves at different directions.} \textbf{(A)} The low frequency part at $\mathrm{k}<0.75k_f$ of the raw $\textit{in situ}$ atomic density and its Fourier transform (upper row) is filtered and the density fluctuations at $\mathrm{k}\geq 0.75k_f$ and its Fourier transform are obtained (lower row). \textbf{(B)} The angular distribution of the Fourier transform magnitude of original atomic density before the filtering (blue open circles) and the corresponding amplitude from fitting the 1D mean density fluctuation $n_\theta(x)$ of the filtered atomic density (orange solid line). The scale of the left and right y axis differ by a factor of 331, which is one half of the area where density waves exist in unit of $\mu m^2$. \textbf{(C)} Three examples of fitting the 1D mean density fluctuation $n_\theta(x)$ at different directions with various Fourier magnitudes (indicated by arrows in \textbf{(B)}.)   }
    \label{FigS2}
\end{figure*}
\section{Phases and amplitudes of density waves}
In order to precisely determine the spatial phase of the density waves at different directions, we develop the following fitting procedure. Since the length scale of the density wave we care about is only around $k_f$, we first filter out the strong low frequency noise below $0.75k_f$ in the Fourier transform of \textit{in situ} density profile $n(x,y)$ and inversely transform it back to obtain the filtered atomic density $\Tilde{n}(x,y)$ as shown in Fig.~\ref{FigS2}A. $\Tilde{n}(x,y)$ is the superposition of plane waves at different directions confined in a finite sized BEC, thus the precision of extracting the phase from its Fourier transform is limited by the small number of density wave periods. In order to avoid this limitation, we first integrate the filtered atomic density along a certain direction $\theta$ normalized by the corresponding integrated circular BEC area to get the averaged 1D density oscillation $n_{\theta}(x) = \int dy n(x,y)/\sqrt{R^2-x^2}$. Then the central part $|x|\leq10~\mu m$ of $n_{\theta}(x)$ is fitted using fit function $f(x) = F_\theta\cos{(k_fx+\phi_\theta)}$, where $F_\theta$ and $\phi_\theta$ are the amplitude and phase of the density wave at $k_f$ and angle $\theta$. Here the step size of angle $\theta$ is chosen to be $1^\circ$ for better resolution compared to the Fourier transform. The amplitude $F_\theta$ and phase $\phi_\theta$ are unaffected by density waves in other directions, which only contribute noise at spatial frequency smaller than $k_f$ or are completely integrated out.    

Fig.~\ref{FigS2}B shows the angular distribution of density wave amplitudes from Fourier transform compared with that from fitting. It can be seen that the results obtained from these two methods are consistent with each other. At the angles indicated by the black arrows in Fig.~\ref{FigS2}B, three examples of the fitting results are shown in Fig.~\ref{FigS2}C. The density oscillation is fit very well when its Fourier amplitude is significant.

\section{Real space pattern recognition algorithm}
We consider each individual $\textit{in situ}$ absorption image as a combination of several common patterns with random orientations and displacements which contribute to the image with different weights. To reveal the common pattern, we align the strongest components from repeated experimental realizations and the weaker ones are averaged to zero. This alignment can be achieved from our real space pattern recognition algorithm.

Here we describe the details of the 2D pattern recognition algorithm (Fig.~3A). We first filter out the low frequency noise at $|\mathbf{k}|<0.75k_f$ from the \textit{in situ} absorption images to get a set of $N = 185$ filtered images of atomic density fluctuations, $\tilde{n}_i(x,y),~i = 1,\cdots,N$ (see Fig.~\ref{FigS2}A). Let $\mathcal{T}_{\theta_i,\mathbf{r}_i}(\tilde{n}_i)$ denote the result of rotating $\tilde{n}_i$ by $\theta_i$ and then translating by $\mathbf{r}_i$, where we impose the constraint $|\mathbf{r}_i|<2\pi/k_f$. The objective function $L$ is the spatial variance of the average image $\bar{n}$ after rotating and translating individual images:

\begin{equation}
    \bar{n}(\{\theta_i\},\{\mathbf{r}_i\}) = \frac{1}{N}\sum\limits_i \mathcal{T}_{\theta_i,\mathbf{r}_i}(\tilde{n}_i),
\end{equation}

\begin{equation}
    L(\{\theta_i\},\{\mathbf{r}_i\}) = \frac{1}{S}\int \bar{n}^2 dxdy - \left(\frac{1}{S}\int \bar{n} dxdy\right)^2,
\end{equation}
where $S$ is the total area of the atomic density fluctuations.
The optimal rotation angles and translation displacements $\{\theta_i\},\{\mathbf{r}_i\}$ are found by maximizing $L$, and the pattern recognized is $\bar{n}$ with the optimal parameters.

Since the rotation angle $\theta_i$ and displacement $\mathbf{r}_i$ are independent degrees of freedom, we perform the optimization of the objective function $L$ in two separate steps. We first find the orientation of each image from the angular distributions of density wave amplitudes $F_\theta$ obtained from fitting (see Fig.~\ref{FigS2}B). The rotation angles $\theta_i$ are changed for individual images in order to maximize the variance of the averaged angular distribution~\cite{Feng2019}. Then the angles are fixed to be the ones after the above optimization before we optimize the displacement of each image. Finally, we translate each image $\tilde{n}_i$ by $\mathbf{r}_i$ to maximize the spatial variance of resulting averaged density fluctuation $\bar{n}$. The recognized common patterns for different modulation schemes are shown in Fig.~3B.

\section{Symmetry decomposition of density patterns}

 We consider each recognized pattern $P$ shown in Fig.~3B as a superposition of normalized two-, four- and six-fold symmetry components $P_{2,4,6}$ with amplitudes $c_{2,4,6}$ and a small offset $c_0$. In order to find the contribution of each symmetry component, we fit the patterns using the following function:

\begin{equation}
    P = c_2 P_2 + c_4 P_4 + c_6 P_6+c_0,
\end{equation}

where
\begin{equation}
    P_2 = \mathcal{R}_{\theta_2}\cos (k_f x+\phi_2),
\end{equation}

\begin{equation}
    P_4 = \frac{1}{\sqrt{2}}\mathcal{R}_{\theta_4}\left[\cos (k_f x + \phi_{4,1})+\cos(k_f y + \phi_{4,2})\right],
\end{equation}

\begin{equation}
\begin{aligned}
    P_6 &= \frac{1}{\sqrt{3}}\mathcal{R}_{\theta_6}\Biggl[\cos (k_f x + \phi_{6,1})\\
    &+ \cos(k_f \left[-\frac{1}{2}x + \frac{\sqrt{3}}{2} y\right] -\frac{1}{2}\phi_{6,1} + \frac{\sqrt{3}}{2}\phi_{6,2})\\
    &+ \cos(k_f \left[-\frac{1}{2}x - \frac{\sqrt{3}}{2} y\right] -\frac{1}{2}\phi_{6,1} - \frac{\sqrt{3}}{2}\phi_{6,2})\Biggr].
\end{aligned}
\end{equation}
Here $\mathcal{R}_\theta[\cdot]$ denotes rotation by angle $\theta$. There are 12 fitting parameters in total: $\{c_2,c_4,c_6\}$ determine the strengths of the symmetry components, $c_0$ determines the overall offset, $\{\theta_2,\theta_4,\theta_6\}$ determine the orientations, and $\{\phi_2,\phi_{4,1},\phi_{4,2},\phi_{6,1},\phi_{6,2}\}$ determine the displacements. The optimal fitting parameters are shown in Table S1.
One example of the symmetry decomposition results for the D$_6$ density pattern under Scheme II is shown in Fig.~\ref{Fig_S4}.
\begin{table}[h!]
    \centering
    \begin{tabular}{|c|c|c|c|c|}
    \hline
         Parameters & units & Scheme I & Scheme II & Scheme III  \\
    \hline
    \hline
        $c_2$ & $\mu m^{-2}$ & 0.302(8) & -1.49(4) & -0.32(1) \\
    \hline
         $c_4$ &$\mu m^{-2}$& -0.080(8) & -0.41(3) & -0.26(1) \\
    \hline
         $c_6$ &$\mu m^{-2}$& 0.070(6) & 1.55(4) & 0.072(8) \\
    \hline
         $c_0$ &$\mu m^{-2}$& -0.009(3) & 0.02(2) & 0.005(4) \\
    \hline
         $\theta_2$ & rad & 1.594(2) & -0.497(2) & 0.274(2) \\
    \hline
         $\theta_4$& rad & 1.537(6) & -0.608(6) & 0.206(2) \\
    \hline
         $\theta_6$ & rad& 1.455(6) & -0.547(1) & 0.008(7) \\
    \hline
         $\phi_2$& rad & 0.96(3) & 5.96(3) & 4.93(3) \\
    \hline
         $\phi_{4,1}$& rad & 4.4(2) & 6.2(1) & -1.72(5) \\
    \hline
         $\phi_{4,2}$& rad & 5.6(1) & 2.5(1) & 3.84(4) \\
    \hline
         $\phi_{6,1}$ & rad& 1.2(1) & 3.43(4) & 2.6(2) \\
    \hline
         $\phi_{6,2}$& rad & 2.7(1) & 5.64(3) & 3.8(2) \\
    \hline
    \end{tabular}
    \vspace{\baselineskip}
    
    Table S1. Optimal fitting parameters for symmetry decomposition.

\end{table}

\begin{figure}[h!]
    \centering
    \includegraphics{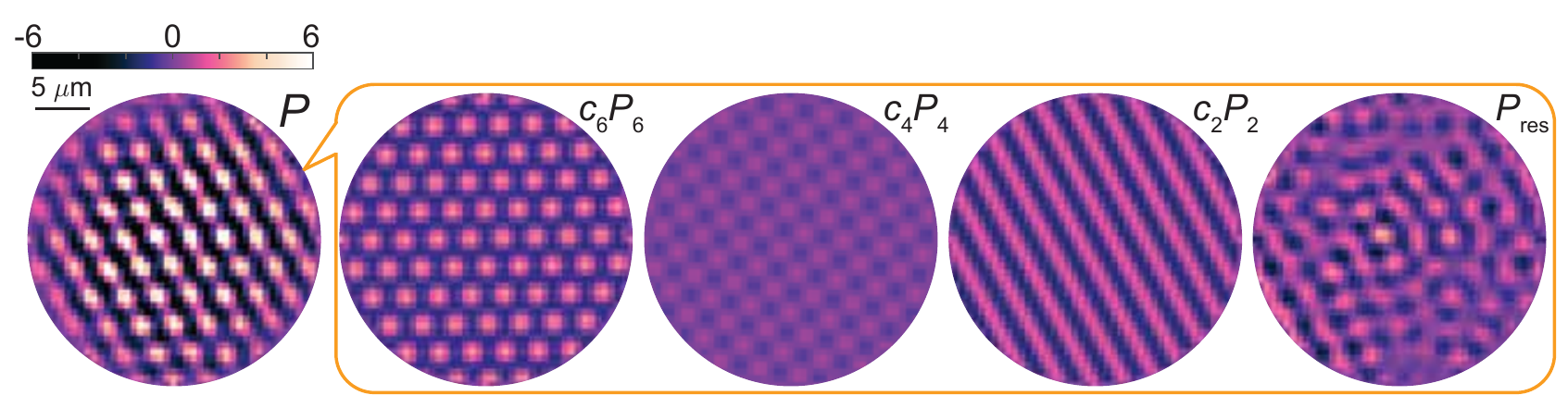}
    \caption{\textbf{Symmetry decomposition of the recognized D$_6$ density pattern under Scheme II.} The pattern $P = c_6P_6+c_4P_4+c_2P_2 + P_{res}$ is projected onto the bases $P_6$, $P_4$ and $P_2$ with weights $c_6$, $c_4$ and $c_2$. The residual $P_{res}$ is dominated by the spatial inhomogeneity of the sample.}
    \label{Fig_S3}
\end{figure}
\section{Hyperbolic growth of D$_6$ pattern}
As we have shown in Fig.~4, for $\mathrm{D}_6$ pattern, only the Fourier modes separated by $2\pi/3$ are coupled together. Since each Fourier mode $A_\theta = \sqrt{N_0}(\hat{a}_{\mathrm{k}}e^{-i\omega t} + \hat{a}^\dagger_{-\mathrm{k}}e^{i\omega t})$ consists of two opposite momentum modes, six momentum modes separated by $\pi/3$ are coupled together. 
Let's first consider a simple model where there are only six modes $\hat{a}_{i},~i = 1,2,\cdots ,6$, separated by $\pi/3$ with momentum $|\mathbf{k}_i| = k_f$. Under driving Scheme II, the equation of motion reads,
\begin{equation}
    \frac{d\hat{a}_i}{dt} = \gamma_1\hat{a}^\dagger_{i+3} + \gamma_2(\hat{a}^\dagger_{i+2}\hat{a}_{i+1}+\hat{a}^\dagger_{i-2}\hat{a}_{i-1}) - \gamma_2^*\hat{a}_{i+1}\hat{a}_{i-1},
    \label{hat_a_hyper}
\end{equation}
where the addition of indices is modulo 6, e.g. $4+3 = 1$.

Here we consider the case where the relative phase $\phi = 0$ between the two frequency components and thus $\gamma_2$ becomes real. After the first 10 cycles of single frequency modulation, the population of each mode is amplified to be larger than the quantum fluctuation. Thus we approximate the operators $\hat{a}_i$ by complex numbers $\Tilde{a}_i$. The equation of motion of amplitude for each mode becomes,
\begin{align}
    \frac{d\Tilde{a}_i}{dt} = \gamma_1\Tilde{a}^*_{i+3}+\gamma_2(\Tilde{a}^*_{i+2}\Tilde{a}_{i+1}+\Tilde{a}^*_{i-2}\Tilde{a}_{i-1}-\Tilde{a}_{i+1}\Tilde{a}_{i-1})
    \label{a_hyper}.
\end{align} 
At the beginning of the two-frequency modulation, we set the population of each mode $n_i(0) = |\Tilde{a}_i(0)|^2$ to satisfy a thermal distribution $p(n) = e^{-n/\bar{n}}/\bar{n}$ with the mean population $\bar{n}$ and $\Tilde{a}_i(0) = \Tilde{a}^*_{i+3}(0)$ with its phase randomly distributed from 0 to $2\pi$~\cite{Hu2019}. Because the growth rates $\gamma_1$ and $\gamma_2$ are real, at any later time t, we always have
\begin{align}
\Tilde{a}_i = \Tilde{a}^*_{i+3}.
\label{a_equal_astar}
\end{align}
Then the Fourier amplitude $A_{\theta_i} = \sqrt{N_0}(\Tilde{a}_ie^{-i\omega t}+\Tilde{a}^*_{i+3}e^{i\omega t}) = 2\sqrt{N_0}\Tilde{a}_i\cos{\omega t}$ and Eq.~\ref{a_hyper} reduces to
\begin{align}
    \frac{d\Tilde{a}_i}{dt} = \gamma_1\Tilde{a}_i + \gamma_2\Tilde{a}_{i+1}\Tilde{a}_{i-1}
    \label{a_reduced}.
\end{align}

Multiplying $\Tilde{a}_i^*$ on both sides of Eq.~\ref{a_reduced} and summing their complex conjugates, we get
\begin{align}
    \frac{d|\Tilde{a}_i|^2}{dt} = 2\gamma_1|\Tilde{a}_i|^2 + 2\gamma_2\Re[\Tilde{a}_i^*\Tilde{a}_{i+1}\Tilde{a}_{i-1}],
    \label{a_i}
\end{align}
where $\Re[\cdot]$ means taking the real part.
Similarly,
\begin{align}
    \frac{d|\Tilde{a}_{i-1}|^2}{dt} &= 2\gamma_1|\Tilde{a}_{i-1}|^2 + 2\gamma_2\Re[\Tilde{a}_{i-1}^*\Tilde{a}_{i}\Tilde{a}_{i-2}],\label{a_i-1}\\
    \frac{d|\Tilde{a}_{i+1}|^2}{dt} &= 2\gamma_1|\Tilde{a}_{i+1}|^2 + 2\gamma_2\Re[\Tilde{a}_{i+1}^*\Tilde{a}_{i+2}\Tilde{a}_i].\label{a_i+1}
\end{align}
Using Eq.~\ref{a_equal_astar}, it can be seen,
\begin{align}
&\Re[\Tilde{a}_i^*\Tilde{a}_{i+1}\Tilde{a}_{i-1}] = \Re[\Tilde{a}_{i+3}\Tilde{a}_{i+1}\Tilde{a}_{i-1}]\nonumber\\
= &\Re[\Tilde{a}_{i-1}^*\Tilde{a}_{i}\Tilde{a}_{i-2}] = \Re[\Tilde{a}_{i+1}^*\Tilde{a}_{i+2}\Tilde{a}_i].
\end{align}
Thus by subtracting two of the equations out of Eqs.~(\ref{a_i}) to (\ref{a_i+1}) and taking the average value on both sides of the equations, we have
\begin{align}
    \frac{dn_{i,i-1}}{dt} &= 2\gamma_1n_{i,i-1},
    \label{n_diff}
\end{align}
where the population difference $n_{i,i-1} = \langle |\Tilde{a}_i|^2\rangle-\langle|\Tilde{a}_{i-1}|^2\rangle$. $n_{i,i+1}$ and $n_{i-1,i+1}$ also satisfy Eq.~\ref{n_diff}. Since at the beginning $|\Tilde{a}_i|^2$, $|\Tilde{a}_{i-1}|^2$ and $|\Tilde{a}_{i+1}|^2$ satisfy the same distribution $p(n)$, they have equal average values $\langle |\Tilde{a}_i(0)|^2\rangle = \langle|\Tilde{a}_{i-1}(0)|^2\rangle = \langle|\Tilde{a}_{i+1}(0)|^2\rangle$, which means the population differences $n_{i,i-1}(0) = n_{i,i+1}(0) = n_{i-1,i+1}(0) = 0$. Thus according to Eq.~\ref{n_diff}, at any later time t, the population differences $n_{i,i-1} = n_{i,i+1} = n_{i-1,i+1} = 0$, i.e.
\begin{align}
    \langle|\Tilde{a}_i|^2\rangle = \langle|\Tilde{a}_{i-1}|^2\rangle = \langle|\Tilde{a}_{i+1}|^2\rangle.
    \label{same_mean}
\end{align}
As is defined in Eq.~4, the three point correlation function at $(\theta,\theta') = (2\pi/3,4\pi/3)$ is,
\begin{align}
    g^{(3/2)}\equiv g^{(3/2)}(\frac{2\pi}{3},\frac{4\pi}{3}) &= \frac{\langle A_{\varphi}A_{\varphi+2\pi/3}A_{\varphi+4\pi/3}\rangle}{\sqrt{\langle |A_{\varphi}|^2\rangle\langle |A_{\varphi+2\pi/3}|^2\rangle\langle |A_{\varphi+4\pi/3}|^2\rangle}}\nonumber\\
    &= \frac{\Re[\langle\Tilde{a}_{i+3}\Tilde{a}_{i+1}\Tilde{a}_{i-1}\rangle]}{\sqrt{\langle|\Tilde{a}_{i+3}|^2\rangle \langle|\Tilde{a}_{i+1}|^2\rangle \langle|\Tilde{a}_{i-1}|^2\rangle}}
    \label{g_threee_half}.
\end{align}
Since the average of the product $ A_\varphi A_{\varphi+\theta}A_{\varphi+\theta'}$ is performed over all the angles with 0$\leq\varphi\leq 2\pi$, it always comes in pair with its complex conjugate, which guarantees that the three point phase correlation function is real. Also, the other possible definitions with one or more of the Fourier amplitudes in $ A_\varphi A_{\varphi+\theta}A_{\varphi+\theta'}$ are equivalent to Eq.~\ref{g_32_main}
 with angular shifts in $\theta$ and $\theta'$, which doesn't show more information.
 Then we take the average value on both sides of Eq.~\ref{a_i} and plug in Eqs.~(\ref{a_equal_astar}), (\ref{same_mean}) and (\ref{g_threee_half}) to get
\begin{align}
    \frac{d\langle |\Tilde{a}_i|^2\rangle}{dt} = 2\gamma_1\langle |\Tilde{a}_i|^2\rangle + 2\gamma_2g^{(3/2)}\langle |\Tilde{a}_i|^2\rangle^{\frac{3}{2}}.
    \label{a_square}
\end{align}
Let's define the root mean square (RMS) of $\Tilde{a}_i$ as $A_{rms} = \sqrt{\langle|\Tilde{a}_i|^2\rangle}$ and plug it into Eq.~\ref{a_square}, we finally arrive at the equation of motion,
\begin{align}
    \frac{dA_{rms}}{dt} = \gamma_1A_{rms} + \gamma_2g^{(3/2)}A_{rms}^2.
    \label{RMS_hyper}
\end{align}
Insert the initial value $A_{rms}(0)$, we obtain the solution of Eq.~\ref{RMS_hyper},
\begin{align}
    A_{rms}(t) = \frac{e^{\gamma_1t}}{1/A_{rms}(0)-\gamma_2\int_0^tg^{(3/2)}(t')e^{\gamma_1t'}dt'}.
\end{align}
This solution exhibits hyperbolic growth that hits a finite time singularity at $t_c$ which satisfies,
\begin{equation}
    \int_0^{t_c}g^{(3/2)}(t')e^{\gamma_1t'}dt' = \frac{1}{\gamma_2A_{rms}(0)}.
\end{equation}
As long as $g^{(3/2)}(t)$ decays slower than $e^{-\gamma_1 t}$, a finite time singularity exists.

\begin{figure}
    \centering
    \includegraphics[width = 86mm]{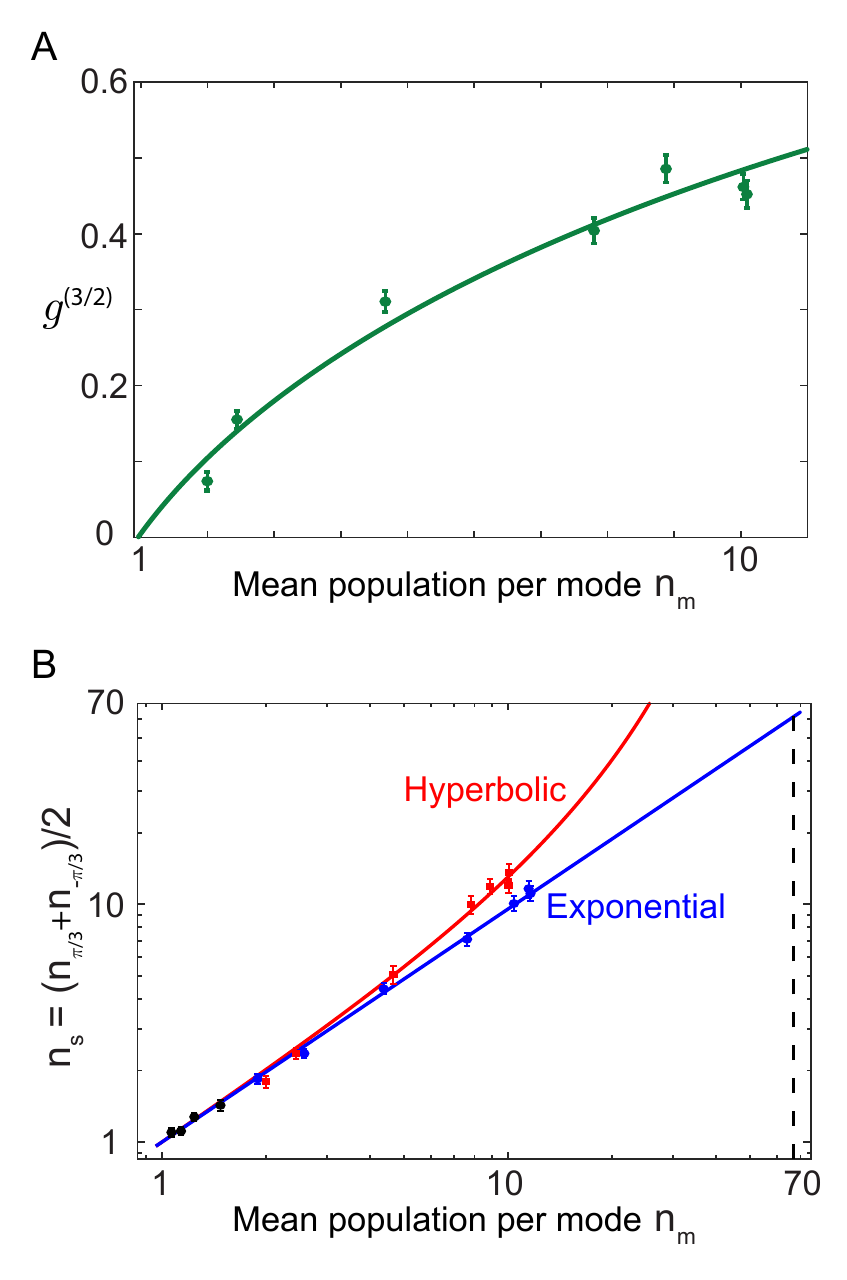}
    \caption{\textbf{The evolution of three point correlation $g^{(3/2)}$ and the mean population at $\pm\pi/3$ relative to the strongest modes during the D$_6$ pattern formation process.} \textbf{(A)} The growth of three point correlation $g^{(3/2)}$ as a function of mean population $n_m$ of modes at all directions in all images. Solid line is the theory curve from fitting using Eq.~\ref{Real_g_three_half}. \textbf{(B)} The growth of mean population $n_s$ of modes at $\pm\pi/3$ relative to the strongest modes versus the mean population per mode $n_m$ under scattering length modulation Scheme II (red squares) compared with that under Scheme I (blue circles). Both the x and y axis are in log scale. The red and blue solid lines are theory curves from fitting using Eq.~\ref{n_pi3_2} and $y = ax$, respectively. The vertical dashed line is the theory prediction when the growth of population $n_s$ diverges during D$_6$ pattern formation.} 
    \label{Fig_S4}
\end{figure}

However, in our experiment, if we look at the mean population $n_m$ of the modes at all directions and images, it doesn't show clear deviation from simple exponential growth. Thus we choose the observable as the mean population $n_s = (n_{\pi/3}+n_{-\pi/3})/2$ at $\pm\pi/3$ relative to the strongest mode in each image. Because the nonlinear coupling between these three adjacent modes, $n_s$ grows faster than $n_m$ and can deviate from exponential growth. If only modulation of single frequency $\omega_1$ is applied, $n_s = n_m$, because they are independent and share the same statistics.

Let's say we always choose $\Tilde{a}_i$ as the strongest mode among all sets of coupled six modes, which have larger fluctuation to begin with. The other two modes $\tilde{a}_{i-1}$ and $\tilde{a}_{i+1}$ at $\pm\pi/3$ relative to it begin with the same mean population as all the other modes with $\langle|\tilde{a}_{i+1}|^2\rangle = \langle|\tilde{a}_{i-1}|^2\rangle$. We model the effect of the strongest sets of modes as an enhancement of $\gamma_2$ by a factor of $\alpha$. Thus the solution of $n_s$ is,

\begin{align}
    n_s^{1/2} = \frac{e^{\gamma_1t}}{1/A_{rms}(0)-\alpha\gamma_2\int_0^tg^{(3/2)}(t')e^{\gamma_1t'}dt'}.
\end{align}
Because the nonlinear term is relatively weak for mean population $n_m$ of all modes, it grows approximately exponentially as
\begin{align}
  n_m = A_{rms}^2(t)\approx A_{rms}^2(0)e^{2\gamma_1t}.  
\end{align}
Then $n_s$ as a function of the mean population $n_m$ of all modes is
\begin{align}
    n_s = \frac{n_m}{\bigg(1-\frac{1}{2}\alpha\epsilon\int_{n_m(0)}^{n_m(t)}g^{(3/2)}(n'_m)/\sqrt{n'_m}dn'_m\bigg)^2},
    \label{n_pi3_1}
\end{align}
where $\epsilon = \gamma_2/\gamma_1$ is the ratio of the two rate constants.

In order to know how $n_s$ grows as a function of $n_m$, we need to determine the evolution of the three point correlation $g^{(3/2)}$ as a function of $n_m$. Combining Eqs.~(\ref{a_reduced}), (\ref{a_i}), 
(\ref{same_mean}) and (\ref{g_threee_half}), we have 
\begin{align}
  \frac{dg^{(3/2)}}{dt} = 3\gamma_2A_{rms}[g^{(2)}-(g^{(3/2)})^2]
  \label{EOM_g_three_half},
\end{align}
where $g^{(2)}$ is the two point correlation function at $\theta = \pi/3$, i.e., $g^{(2)} = \langle |A_{\varphi}|^2|A_{\varphi+\pi/3}|^2\rangle/\langle|A_{\varphi}|^2\rangle^2 = (\langle|\Tilde{a}_{i+1}|^2|\Tilde{a}_{i}|^2\rangle + \langle|\Tilde{a}_{i}|^2|\Tilde{a}_{i-1}|^2\rangle + \langle|\Tilde{a}_{i+1}|^2|\Tilde{a}_{i-1}|^2\rangle )/3\langle|\Tilde{a}_i|^2\rangle^2$. In the perturbation regime where the population of modes in directions separated by $\pi/3$ are almost uncorrelated,~i.e.~$g^{(2)}\approx 1$, the three point phase correlation is given by
\begin{align}
    g^{(3/2)} = 1-\frac{2}{1+\exp[6\epsilon(\sqrt{n_m}-A_{rms}(0))]}
    \label{Real_g_three_half}.
\end{align}
Inserting the above result into Eq.~\ref{n_pi3_1}, we arrive at
\begin{align}
    n_s = n_m[1-\alpha\epsilon(A_{rms}(0)-\sqrt{n_m})+\frac{\alpha}{3}\ln{(1-g^{(3/2)})}]^{-2}.
    \label{n_pi3_2}
\end{align}
Since our model only considered 6 excited modes, here the mean population $n_m$ of a single mode is $1/6$ of the total mean population. In our experiment, the total number of excited modes at $|\mathbf{k}| = k_f$ is $N_{mod} = 1.62/Rk_f \approx 136$~\cite{Clark2017}. In order to generalize Eq.~\ref{Real_g_three_half} and Eq.~\ref{n_pi3_2} for multiple sets of 6 modes with $\pi/3$ angular spacing, we need to do the replacements: $n_m\rightarrow\frac{N_{mod}}{6}n_m$, $A_{rms}(0)\rightarrow\sqrt{\frac{N_{mod}}{6}}A_{rms}(0)$ and $n_s\rightarrow\frac{N_{mod}}{6}n_s$. This is equivalent to replace $\epsilon$ by $\sqrt{\frac{N_{mod}}{6}}\epsilon = \sqrt{\frac{N_{mod}}{6}}\frac{\gamma_2}{\gamma_1}$.

Using Eq.~\ref{Real_g_three_half} to fit the data with $\epsilon$ and $A_{rms}(0)$ as fitting parameters as shown in Fig.~\ref{Fig_S4}A, we get $\epsilon = 0.08$ and $A_{rms}(0) = 0.98$. Thus $\gamma_2/\gamma_1 = 0.01$, which is consistent with the experimental value 0.003. The discrepancy is attributed to the exclusion of other collision processes that are also involved in the experiment, such as the pair generation from BEC at $|\mathbf{k}| = k_f/\sqrt{2}$ and secondary collision processes that lead to D$_4$ pattern.
Then we use the value of $\epsilon$ and $A_{rms}(0)$ from fitting the three point correlation function $g^{(3/2)}$ and set $\alpha$ as another fitting parameter to fit $n_s$ versus $n_m$ as shown by the red solid line in Fig.~\ref{Fig_S4}B, which gives $\alpha = 2.78$. On the other hand, for single frequency modulation under Scheme I, we use the fit function $y = ax$ and the best fit is obtained with $a = 0.98$ as shown by the blue solid line in Fig.~\ref{Fig_S4}B.

\section{Evolution of the phase relation of modes forming D$_6$ density wave pattern}
In order to study how the phase relation of modes that form hexagonal lattices evolve from completely uncorrelated to concentrated around the plane $\phi_0 + \phi_{2\pi/3}+\phi_{4\pi/3}$ = 0, we perform numerical calculation based on Eq.~\ref{a_hyper}. Here we consider BEC with depletion, which couple to multiple sets of 6 modes with $\pi/3$ angular spacing at the same time. The corresponding equations of motion are:
\begin{align}
    \frac{d\tilde{a}_i}{dt} = [\gamma'_1N_0(t)-\gamma_e]\tilde{a}^\dagger_{i+3}+\gamma'_2&\sqrt{N_0(t)}(\tilde{a}^\dagger_{i+2}\tilde{a}_{i+1}+\tilde{a}^\dagger_{i-2}\tilde{a}_{i-1}-\tilde{a}_{i+1}\tilde{a}_{i-1})\\
    N_0(t) &= N_0 - \sum_{i = 1}^{N_{mod}}|\tilde{a}_i|^2,
\end{align}
where the growth rates $\gamma'_1 = \gamma_1/N_0$ and $\gamma'_2 = \gamma_2/\sqrt{N_0}$. The decay rate due to modes flying out of the condensate is $\gamma_e\sim v/R$, where the velocity of the modes $v = \hbar k_f/m$ and $R$ is the radius of the condensate.  

\begin{figure}
    \centering
    \includegraphics[width = 86mm]{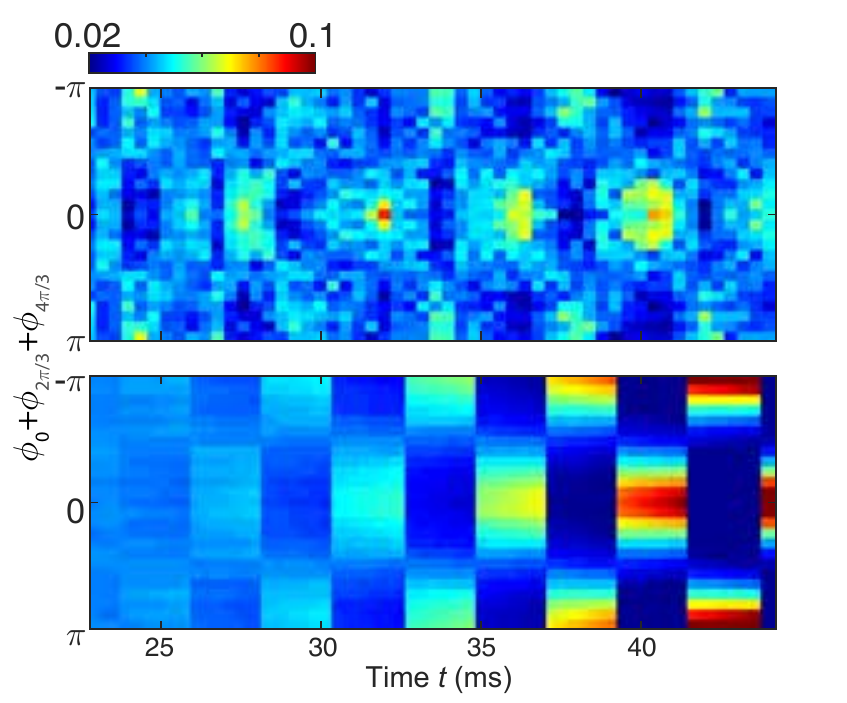}
    \caption{\textbf{Evolution of the phase distribution of $\phi_0+\phi_{2\pi/3}+\phi_{4\pi/3}$ during  D$_6$ density wave pattern formation process.} The upper panel is from the experiment under Scheme II where we perform \textit{in situ} imaging of the condensate at different times. The peak position of the phase distribution oscillates between 0 and $\pi$ and gets more concentrated as time evolves. The lower panel is from the numerical calculation with the modulation amplitudes $a_1 = 22.5~a_0$ and $a_2 = 63.5~a_0$ for frequency components of 450~Hz and 225~Hz, respectively. The escaping rate of momentum modes is 39~Hz. Other parameters are the same as the experiment.  }
    \label{Fig_S5}
\end{figure}

The simulation starts from the beginning of the second pattern forming stage. At the end of the first seeding stage, the population n in each mode $\tilde{a}_i$ is thermally distributed according to the probability distribution $p(n) = e^{-n/\bar{n}}/\bar{n}$ with the mean population $\bar{n} = 2$. The phase of each mode $\tilde{a}_i$ is uniformly distributed from 0 to $2\pi$ and the modes in opposite directions are correlated as $\tilde{a}_i = \tilde{a}_{i+3}^*$. The simulation is repeated for 5000 times and each time the initial conditions of the phase and amplitude are independently sampled from their distributions. We finally take the phase of $ \tilde{a}_ie^{-i\omega t}+\tilde{a}_{i+3}^*e^{i\omega t}$ as the phase of Fourier modes in the lab frame. The amplitude of scattering length modulation $a_1$ and $a_2$ and the escape rate $\gamma_e$ are chosen as fitting parameters while all the other parameters are the same as our experiment for Scheme II. The green line in Fig.~\ref{Fig4}F is the result after 22.4~ms evolution time, using the initial condition of mean population at 22.6~ms in our experiment. The corresponding amplitudes of modulation are $a_1 = 24~a_0$, $a_2 = 68.5~a_0$ and $\gamma_e = 39~\mathrm{Hz}$. The evolution of the phase distribution of $\phi_0+\phi_{2\pi/3}+\phi_{4\pi/3}$ within individual Floquet periods is also calculated as shown in the lower panel of Fig.~\ref{Fig_S5}, which is consistent with the experimental result in the upper panel. The peak position $\phi_{peak}$ of the phase distribution oscillates between 0 and $\pi$, due to the standing wave nature of the density waves. This also means the real space pattern changes back and forth between hexagonal lattice ($\phi_{peak} = 0$) and honeycomb lattice ($\phi_{peak} = \pi$). However, in the rotating frame, the phase distribution is always centered at 0, thus the three point correlation $g^{(3/2)}$ is always positve. This ensures the hyperbolic growth since the second term in Eq.~5 is positive.

\end{document}